\documentclass[%
 reprint,
 amsmath,amssymb,
 aps,
]{revtex4-2}

\usepackage[latin1]{inputenc}
\usepackage[pdftex]{graphicx}
\usepackage{epsfig}
\usepackage{graphicx}
\usepackage{dcolumn}
\usepackage{bm}
\usepackage{color}
\usepackage{natbib}
\newcommand {\be}{\begin{equation}}
\newcommand {\ee}{\end{equation}}
\newcommand {\ba}{\begin{eqnarray}}
\newcommand {\ea}{\end{eqnarray}}

\usepackage{amsmath}  
\usepackage{amsfonts} 
\usepackage{graphicx} 
\usepackage{nicefrac}
\usepackage{hyperref}
\usepackage{ulem}

\begin{document}


\title{Performance at maximum figure of merit  for a Brownian Carnot refrigerator} 

\author{O. Contreras-Vergara${^1}$}
\author{G. Valencia-Ortega${^2}$}
\author{N. S\'anchez-Salas${^1}$}\email{nsanchezs@ipn.mx}
\author{J. I. Jim\'enez-Aquino${^3}$} 
\affiliation{${^1}$Departamento de F\'isica,
Escuela Superior de F\'isica y Matem\'aticas, Instituto Polit\'ecnico Nacional, 
 Edif. 9 UP Zacatenco, CP 07738, CDMX, M\'exico.}
\affiliation{${^2}$Divisi\'on de Matem\'aticas e Ingenier\'ia, Facultad de Estudios Superiores Acatl\'an, Universidad Nacional Aut\'onoma de M\'exico, Av. Alcanfores y San Juan Totoltepec, Santa Cruz Acatl\'an, Naucalpan de Ju\'arez, 53150, Estado de M\'exico, M\'exico.}
\affiliation{${^3}$Departamento de F\'isica, Universidad Aut\'onoma Metropolitana-Iztapalapa,
 C.P. 09340, CDMX, M\'exico.}

 \date{\today}

\begin{abstract}
This paper focuses on the coefficient of performance (COP) at maximum figure of merit $\chi$ for a Brownian Carnot-like refrigerator, within the context of symmetric Low-Dissipation approach. Our proposal is based on the Langevin equation for a Brownian particle bounded to a harmonic potential trap, which can perform Carnot-like cycles at finite time. We show that under quasistatic conditions the COP has the same expression as the macroscopic Carnot refrigerator. However, for irreversible cycles at finite time and under symmetric dissipation, the optimal COP is the counterpart of Curzon-Ahlborn efficiency for irreversible macroscopic refrigerators.       

\end{abstract}
\maketitle 

\section{Introduction} 

Recently the study of the efficiency at maximum power for 
stochastic heat engine performing Carnot-, Stirling- and Ericsson-like cycles at finite time, has been reported \cite{Contreras2023}. This  was inspired by the work related to 
macroscopic heat engines performing finite-time Carnot cycles, and operating under low-dissipation conditions. In 2010, this approach  was first established for a macroscopic Carnot-like engine which can operate under irreversible conditions at finite-time \cite{Esposito10}. Under these non-equilibrium conditions the dissipative processes come directly from the heat exchange between the system and the thermal reservoirs. In several studies  it has been shown that endoreversible and exoreversible cycle models can agree with low-dissipation approach, by considering appropriate constraints \cite{Schmiedl2007,DeTomas12,Wang12,Apertet2013,Holubec2016,Gonzalez2016,Rana2016,Ma2018,Johal2019,Ma2021,Chen2022,Zhao2022}. Two years later, low-dissipation approach could be successfully extended to study the optimal COP at maximum figure of merit of macroscopic Carnot-like refrigerators, under symmetric \cite{DeTomas12} and symmetric-asymmetric  
\cite{Wang12} low-dissipation conditions. In the later the corresponding bounds for the COP 
were established.

In Ref. \cite{DeTomas12}, the authors proposed a unified optimization criterion for Carnot engines and refrigerators, which consists in maximizing the figure of merit, defined as $\chi=z Q_{in}/t_{cycle}$, where $z$ is the converter efficiency, $Q_{in}$ the heat absorbed by the working system, and $t_{cycle}$ the cycle time. When the criterion is applied to Carnot-like engines, $Q_{in}=Q_h$ is the heat absorbed from the hot reservoir, $z=\eta=-W/Q_h$, is the efficiency of conversion, and thus the figure of merit reads $\chi^{(E)}=-W/t_{cycle}$, which is consistent with the power output of the heat engine. In the case of a Carnot-like refrigerator 
$Q_{in}=Q_c$ is the heat exchanged with the cold reservoir, $z=\varepsilon=Q_c/W$ the COP, and $W$ the amount of input work. The figure of merit becomes $\chi^R=\varepsilon Q_c/t_{cycle}$. In summary, in the case of Carnot-like engines, the authors showed that, under symmetric low-dissipation condition, the efficiency at maximum power coincides with the Curzon-Ahlborn efficiency, 
$\eta_{_{CA}} =1 - \sqrt{T_c/T_h}$, 
where $T_h$ and $T_c$ are the temperatures of the hot and cold reservoirs, respectively \cite{Curzon1975,Esposito10}. For refrigerators the criterion provides
also, under symmetric low-dissipation condition the Curzon-Ahlborn COP  given by $\varepsilon_{_{CA}} =[1/\sqrt{1- T_c/T_h}]-1$, which was first derived and reported in \cite{Yan90} for an endoreversible Carnot-type refrigerator.
On the other side, in Ref. \cite{Wang12} the optimal COP ($\varepsilon_*$), has  been calculated for both asymmetric and symmetric low-dissipation schemes. Likewise, it is also shown that $\varepsilon_*$ is bounded in the interval $0\leq\varepsilon_*\leq \sqrt{9+8\varepsilon_c}-3)/2$, where $\varepsilon_c=T_c/(T_h-T_c)$ is the Carnot coefficient of performance for reversible refrigerators.

In the present contribution, we follow the study reported in \cite{Contreras2023} and  \cite{Wang12}  to extend the asymmetric low-dissipation approach to the case of Brownian Carnot-like refrigerators. We assume that this system consists on a Brownian particle confined in an optical trap (represented by a harmonic potential), performing finite-time Carnot-like cycles between two thermal baths at temperature $T_h$ and $T_c$, which can be tuned via the internal noise intensity. It must be highlighted that within the low-dissipation approach, the adiabatic processes in a macroscopic Carnot heat engine are considered instantaneous, and the irreversible effects are present only in the two isothermal processes. This assumption also has been considered in stochastic Carnot-like heat engines, which means that the Brownian particle relaxation time (time to reach the equilibrium state) is much faster than the quenching time of the temperature \cite{Contreras2023}. In fact, it has been pointed out that in the experiment at small scales it is very difficult to keep  hot and cold reservoirs thermally isolated, so rather than coupling the colloidal
particle periodically to different heat baths, the temperature of the surrounding liquid is suddenly changed \cite{Blickle12}.\\ 
Our theoretical analysis again relies upon the overdamped Langevin equation associated with a Brownian particle bounded in a time-dependent harmonic trap potential. Both the potential stiffness $\kappa(t)$ and temperature $T(t)$ of the surroundings are time-dependent parameters which can externally be controlled \cite{Martinez16}. Thus, for a stochastic Carnot-like refrigerator, we also are interested in calculating the corresponding thermodynamic parameters but as average quantities. That is, the COP is defined as $\langle\varepsilon\rangle=\langle Q_c\rangle/ \langle W\rangle$, being $\langle W \rangle =\langle Q_h\rangle-\langle Q_c\rangle$ the amount of input work. Then, the figure of merit must $\langle\chi\rangle=\langle \varepsilon\rangle\langle Q_c\rangle/t_{cycle}$, being again $t_{cycle}$ the cycle time. In this work, we are interested in obtaining the COP at maximum figure of merit under symmetric low-dissipation approach, and showing that again the COP for the Curzon-Ahlborn refrigerator can be written in analytical form. In a similar way as done in \cite{Contreras2023}, and according to low-dissipation approach, the starting point is a reversible Brownian Carnot-like refrigerator with infinite cycle time, and then the entropy change in a cycle must be zero. In the case of an irreversible Brownian Carnot-like refrigerator, there is an entropy production in each isothermal process, which is assumed inversely proportional to the time required for performing such process. 
The strategy is also based on the transformation of Langevin equation into a macroscopic one for the average value $\langle x^2(t)\rangle$, which in the long time limit, plays the role of a state-like equation. This allows us to obtain all the thermodynamic quantities under quasistatic conditions, and the irreversible effects are taken into account using the low-dissipation approach.

\section{Brownian Carnot-like refrigerator}

As in the case of a stochastic heat engine \cite{Contreras2023}, a Brownian particle is bounded to a harmonic trap potential 
with time-dependent stiffness and embedded in a heat bath of time-dependent temperature. The macroscopic overdamped Langevin equation can read as   
\begin{equation}
    \gamma  \frac{d \langle x^2\rangle}{dt}=-2\kappa(t) \langle x^2\rangle + 2k_{_B}T(t),  \label{mex2a}
\end{equation}
where $\gamma$ is the friction coefficient and $k_{_B}$ the Boltzmann's constant. Instead of straightforward solving Eq. (\ref{mex2a}), we can take advantage from low-dissipation strategy for a Carnot-like refrigerator as in \cite{DeTomas12,Wang12}, and thus we calculate the nonequilibrium quantities around the equilibrium state. The corresponding equilibrium quantities can be calculated using the state-like equation associated with $\langle x^2\rangle_{st}$. 

\subsection{Quasistatic Brownian Carnot-like refrigerator} 

A Carnot-like refrigerator is a Carnot cycle operates in  the opposite direction. This refrigerator extracts a certain amount of heat from the cold bath, requiring a certain amount of work input, and delivering a certain amount of heat to the hot bath. In this regard, the Brownian heat engine proposed in Fig. 1 of Ref. \cite{Contreras2023} becomes a Brownian refrigerator. That is, the optical trap transfers energy in the form of external work to the Brownian particle against the temperature gradient (from the ``cold'' bath to the ``hot'' one). It is expected to obtain a process that can be called Brownian Carnot-like refrigerator. Thus the COP for this type of stochastic devices is defined as $\langle \varepsilon_c\rangle \equiv \langle Q_c\rangle /\langle W\rangle =\langle Q_c \rangle/(\langle Q_h\rangle-\langle Q_c\rangle)$, where $\langle W\rangle =\langle Q_h\rangle -\langle Q_c\rangle$ is the average of the amount of input works in the cycle. In the equilibrium state the stiffness as well as the temperature become constants and the state-like equation reads $\langle x^2\rangle_{eq}=k_{_B}T/{\kappa}$. Then, a state-point is characterized by $(\langle x^2\rangle, \kappa, T)$ as thermodynamic variables \cite{Contreras2023}. From now on, we will define the average of any microscopic quantity as $\langle y\rangle\equiv y$, for example $\langle W\rangle\equiv W$, $\langle Q\rangle\equiv Q$, etc.\\ According to \cite{Contreras2023}, the total amount of energy that the system (the particle) can exchange with its surroundings is $E={1\over 2} \kappa \langle x^2\rangle + {1\over 2}k_{_B}T$
and thus $E=k_{_B}T$. 
\begin{figure}
    \centering
    \includegraphics[scale=0.2]{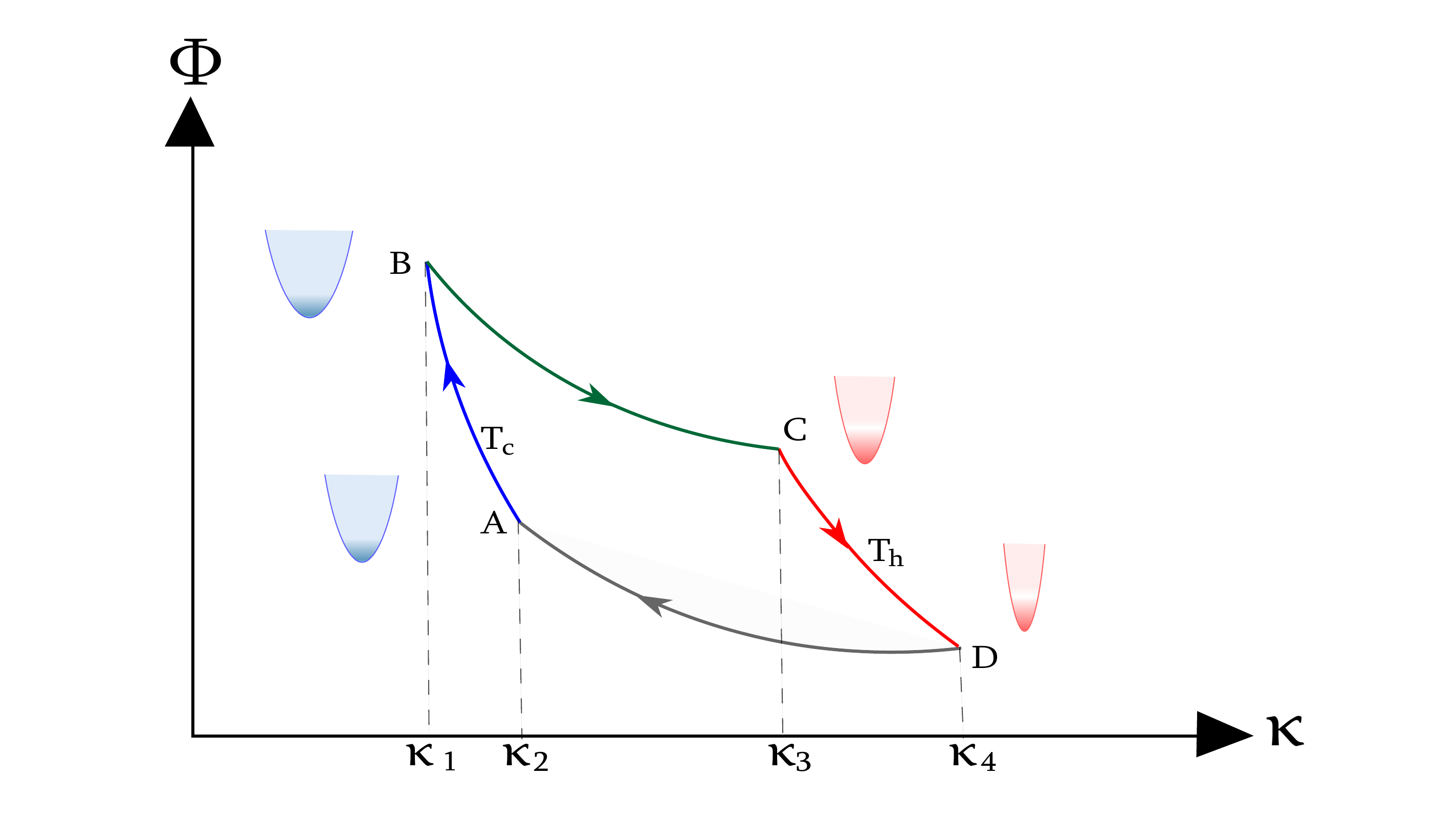}
    \caption{$\Phi$-$\kappa$  cycle of a Brownian Carnot-like refrigerator where (i) an isothermal expansion (A-B path); (ii) an adiabatic compression (B-C path); (iii) an isothermal compression (C-D path) and (iv) an adiabatic expansion (D-A path). }
    \label{Carnot}
\end{figure}
For this system, all the thermodynamic quantities can be calculated along the quasistatic trajectories of a Carnot-like cycle, with  the total energy equation and the first law-like of thermodynamics given by  $dE=d^{\prime}Q+d^{\prime}W$, where $d^{\prime}W={1\over 2}x^2 d\kappa$ and 
$d^{\prime}Q={1\over 2}\kappa d\langle x^2\rangle + {1\over 2}k_{_B} dT$. The entropy is $S=(k_{_B}/2)[\ln (2 \pi k_{_B}T /\kappa)+1]$, the same as the one given in Sec. IV of Ref. \cite{Contreras2023}. The auxiliary conjugate thermodynamic variable  $\Phi$, also satisfies the state-like equation   
$\Phi=k_{_B}T/2\kappa=\langle x^2\rangle/2$. It must be noticed that the parameters $\Phi$ and $\kappa$ play the role of a pressure $p$ and volume $V$ for an ideal gas,  as in classical Thermodynamics. 
So, the total work $W$ and exchange heat $Q$ with the surroundings along a quasistatic trajectory, from a  state $a$ to another state $b$, are given by
\be
W_{ab} = {1\over 2}\int _a^b \langle x^2\rangle \, d\kappa , \quad
Q_{ab}={1\over 2}\int_a^b  \kappa d\langle x^2\rangle + {1\over 2}k_{_B} (T_b -T_a), \label{WQab}
\ee 
The ideal Carnot cycle involves two quasistatic isothermal processes ($T$ is constant and $\kappa$ changes), and two 
reversible adiabatic processes ($\kappa$ and $T$ change  along the adiabatic path, with $\kappa ={\rm const} \, T^2$ \cite{Contreras2023,Martinez2015}). In Fig. \ref{Carnot}, a Brownian Carnot-like refrigerator operating in cycles is sketched, it goes in the following way: (i) {\it isothermal expansion} from $A\to B$, (ii) {\it adiabatic compression} from $B\to C$, (iii) {\it isothermal compression} from $C\to D$, and finally (iv) {\it adiabatic expansion} from $D\to A$. We can calculate the COP by means of the input work and heat that the system can exchange through the ishotermal  processes, taking also into account that from the adiabatic equation it satisfies that $\kappa_3/ \kappa_1=\kappa_4/\kappa_2$.
The calculations are similar to those reported in \cite{Contreras2023}, and therefore, the COP of a Brownian Carnot-like refrigerator becomes
\be
\varepsilon_c={Q_c\over Q_h-Q_c}=\frac{k_{_B}T_c \ln\left({\kappa_2\over \kappa_1}\right)}{k_{_B}\left[T_h\ln\left({\kappa_4\over \kappa_3}\right) - T_c\ln\left({\kappa_2\over \kappa_1}\right)\right]}, \label{copa}
\ee
and therefore  $\varepsilon_c=T_c/(T_h-T_c)$. It coincides with the COP obtained from the  classical thermodynamics which only depends on the temperatures of the thermal baths. 

\subsection {Irreversible Brownian Carnot-like refrigerator} 

If the cyclic processes are no longer reversible but irreversible at finite-time, then the dissipative processes play an important role. In this case, it has been shown that a suitable theoretical approach used to characterize out of equilibrium macroscopic heat engines and refrigerators, is the low-dissipation approach \cite{DeTomas12, Wang12}. To study the model of a Carnot-like cycle at finite time for a Brownian refrigerator, we adopt the similar idea proposed in \cite{Wang12} for a Carnot-like engine, that is: \\
i) {\it Isothermal expansion:} The cycle begins when the Brownian particle (working substance) is in contact with a ``cold'' thermal bath at constant temperature $T_c$; during the time interval $0<t<t_c$, the expansion process means that the control parameter decreases from $\kappa_2 \to \kappa_1(<\kappa_2)$, while $T(t)=T_c$. In this finite process an amount of heat $Q_c$ is absorbed by the particle ( it is assumed that $Q_c>0$ and $Q_h<0$). In this isothermal process the variation of entropy can be written as 
\be
\Delta S_c= {Q_c\over T_c} + \Delta S_c^{ir} , 
\label{dsc} \ee
where $\Delta S_c^{ir} \ge 0$ is  one part of the entropy production  and fulfills $\Delta S_c\ge \Delta S_c^{ir}$\\
ii) {\it Adiabatic compression:} In a similar way as done with the Brownian Carnot-like heat engine, this adiabatic process occurs instantaneously, the particle suddenly decouples from the ``cold'' thermal bath at $T_c$ and then comes into contact with the ``hot''  one at $T_h$. The compression means that during this transition process the potential stiffness suddenly is switched from $\kappa_1 \to \kappa_3 (>\kappa_1)$. This physically means that the relaxation time (time to reach the equilibrium state) of Brownian particles is much faster with  respect to the quenching time of the temperature. In this sudden adiabatic compression $Q_2=0$ and thus the entropy change $\Delta S_2=0$.\\
iii) {\it Isothermal compression:}
In this process, the Brownian particle is in contact with the ``hot'' thermal bath at temperature $T_h$, and the potential stiffness again is switched from  
$\kappa_3 \to \kappa_4$, while $T(t)=T_h$ for $t_c<t<t_c+t_h$. In this finite process an amount of heat $Q_h$ is released by the particle to the hot bath. The variation of entropy in this process is now 
\be
\Delta S_h= -{Q_h\over T_h} + \Delta S_h^{ir} , 
\label{dsh} \ee
and $\Delta S_h^{ir}\ge 0$ is the other part of the entropy production. \\
iv) {\it Adiabatic expansion:} In this last branch, the Brownian particle again suddenly decouples from the ``hot'' thermal bath at $T_h$ and then comes into contact with the ``cold'' one at $T_c$.  During this transition, the potential stiffness is switched from $\kappa_4\to \kappa_2(<\kappa_4)$. In this branch, $Q_4=0$ and $\Delta S_4=0$. \\

\section{COP at maximum figure of merit}

The total change of entropy vanishes in the whole cycle, and thus the change of entropy in the two isothermal processes  fulfill $\Delta S_c=-\Delta S_h\equiv \Delta S>0$. Using Eqs. (\ref{dsc}) and (\ref{dsh}) the figure of merit reads
 \be
 \chi={T_c^2(\Delta S-\Delta S_c^{ir})^2\over (T_h-T_c)\Delta S + T_c\Delta S_c^{ir}+T_h\Delta S_h^{ir}](t_c+t_h)} , \label{chi}
 \ee
where $\Delta S=(k_{_B}/2) \ln(\kappa_1/\kappa_2)$. Because of the mathematical form of function $\chi$, its maximum value is reached when $\Delta S_c^{ir}$ and $\Delta S_h^{ir}$ fulfill a minimum value with respect to the protocols $\kappa_c(t)$ and $\kappa_h(t)$. Within Low-Dissipation approach,  $\Delta S_c^{ir} \propto L_c(t_c)$ and $\Delta S_h^{ir} \propto L_h(t_h)$, which  are monotonous decreasing functions of $t_c$ and $t_h$, respectively. That is $L_c(t_c)\sim L_c/t_c$ and $L_h(t_h)\sim L_h/t_h$. This means, when times $t_c\to\infty$ and $t_h\to\infty$, the corresponding entropy production terms tend to zero. Then, we have that  
\ba  
Q_c=T_c[\Delta S- L_c x_c] , \label{qc} \\
Q_h=T_h[\Delta S+L_h x_h] , \label{qh} \ea
where $x_c=1/t_c$ and $x_h=1/t_h$, and thus the COP for Brownian refrigerators becomes 
\be
\varepsilon={Q_c\over Q_h-Q_c}= {T_c(\Delta S-L_c)\over (T_h-T_c)\Delta S + T_cL_c+T_hL_h} . \label{copb}  
\ee
We now proceed to calculate the COP at maximum figure of merit. First of all the figure of merit given by Eq. (\ref{chi}), can be written as  
\be
\chi = \frac{Q_c^2 x_c x_h}{Q_hx_h +Q_hx_c - Q_cx_h - Q_cx_c} .  \label{chia}
\ee
The optimization criterion leads us to calculate ${\partial L\over\partial x_{h,c}}=0$.  And so, with respect to variables $x_c$ and $x_h$, the following two equations arise:  
\ba
x_h(Q_h - Q_c) &=& \left(\frac{2Q_h}{Q_c} - 1\right)x_cT_cL_c^{\prime}(x_h + x_c) , \label{opt1} \\ 
\nonumber\\
x_c(Q_h - Q_c)&=& T_hL_h^{\prime}x_h(x_h+x_c) ,
 \label{opt2} \ea
where $L^{\prime}_c$ and $L^{\prime}_h$ are the derivative of {$L_c$ and $L_h$ respect to $x_c$ and $x_h$. Dividing Eqs. (\ref{opt1}) and (\ref{opt2}), and  taking into account Eqs. (\ref{qc})-(\ref{copb}), we derive the COP $\varepsilon_*$, at $\chi$-maximum figure of merit   
\be
 \epsilon_*T_h L_h^{\prime} x_h^2=(\epsilon_* +2)T_c L_c^{\prime} x_c^2 .  \label{opta} 
\ee
On the other hand, adding Eqs. (\ref{opt1}) and (\ref{opt2}), it is possible to show that 
\be
{1\over \epsilon_*} ={T_h-T_c \over T_c}+ {T_h( L_h+L_c) \over  2 T_c  L_c^{\prime} x_c  +\epsilon_* T_h L_h^{\prime} x_h  +\epsilon_* T_c L_c^{\prime} x_c }  , \label{optb}
\ee
which can be rewritten as 
\be
{1\over \epsilon_*} ={1\over\epsilon_c}+ {1+\varepsilon_c\over N\varepsilon_* (1+\varepsilon_c)+(2\varepsilon_c-\varepsilon_*)M }  , \label{optc}
\ee
where $
M={L_c^{\prime} x_c\over L_c+L_h}$, and $N={L_c^{\prime}x_c+L^{\prime}_hx_h\over L_c+L_h}$. We also assume that if $L^{\prime}_c=\Sigma_c$ and $L^{\prime}_h=\Sigma_h$ are two dissipation constants, as proposed by Wang \cite{Wang12}, thus $N=1$ and 
$M={\Sigma_c x_c\over \Sigma_h x_h+\Sigma_c x_c}$. Moreover, in the symmetric  case 
$\Sigma_c=\Sigma_h=\Sigma$  \cite{DeTomas12}, and making use of Eqs. (\ref{opta}) and (\ref{optc}), it can be shown that
\be
 (z-1)(2z-1)-(1+\epsilon_c)
 =  \sqrt{(1+\epsilon_c)(2z+\epsilon_c)} . \label{z}   \ee
with $z={\varepsilon_c\over\varepsilon_*}$. Following the algebraic steps, we arrive to the COP at maximum figure of merit  
\be
\epsilon_*\equiv\epsilon_{_{CA}}
 =\sqrt{1+\epsilon_c} -1 = {1\over\sqrt{1-\theta }}-1,  \label{cap}  
\ee
where $\theta=T_c/T_h$. This COP  is precisely the counterpart of Curzon-Ahlborn efficiency for refrigerators. This result was first derived by Yan and Chen for the particular case of an
endoreversible Carnot-type refrigerator \cite{Yan90}. Also, making use of Eqs. (\ref{qc}), (\ref{qh}) (\ref{opt1}) and (\ref{opt2}), we can show that critical value of times $t^*_c$ and $t^*_h$ are  
\ba
t^*_c&=&{2\Sigma\over \Delta S}\left(1+ {1\over \sqrt{1-\theta}} \right) \cr\cr
&=&{4\Sigma\over k_{_B}\ln(\kappa_1/\kappa_2)}\left(1+ {1\over \sqrt{1-\theta}} \right) , \label{tcm} \\ 
t^*_h&=&{2\Sigma\over \Delta S}\left({1\over \sqrt{1-\theta}} \right)\cr\cr
&=&{4\Sigma\over k_{_B}\ln(\kappa_1/\kappa_2)}\left({1\over \sqrt{1-\theta}} \right) , \label{thm}
\ea
both was obtained in \cite{DeTomas12} for macroscopic refrigerators. In fig. \ref{grafCAref}, numerical evaluations, generated by the authors in \cite{qi2022}, show that the trend of the values for the COP of a Brownian refrigerator model tend to $\varepsilon_{_{CA}}$ as $\theta$ tends to 1, showing the robustness of low-dissipation approach.
\begin{figure}
    \centering
    \includegraphics[scale=0.8]{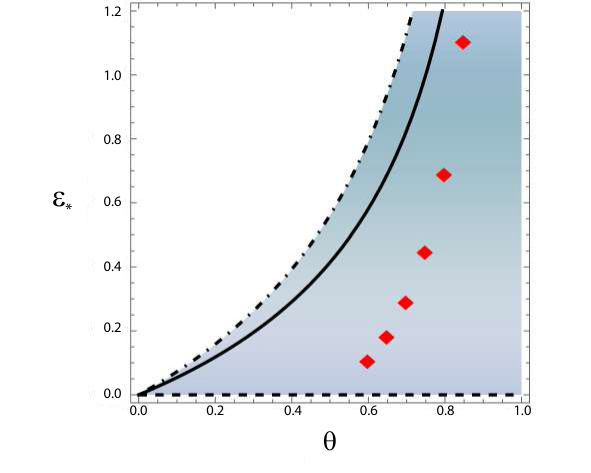}
    \caption{Plot of COP ($\varepsilon_{*}$) at maximum $\chi$-figure of merit of an irreversible Carnot-like refrigerator as a function of $\theta$. The Curzon-Ahlborn COP is denoted by the solid line. Diamonds represent the numerical evaluations in \cite{qi2022}. The upper and lower boundes for the asymmetric case are marked by a dot dashed line and a dashed line, respectively.}
    \label{grafCAref}
\end{figure}
In conclusion, under quasistatic condition it has been shown that if the Brownian Carnot-like heat engine operates in opposite direction, and with the purpose of extracting an amount of heat from the thermal bath at $T_c$; a kind of stochastic Carnot-like refrigerator can be obtained. In this case, the COP is the same as the macroscopic Carnot refrigerator, as expected. In the case of Brownian Carnot-like refrigerator, we have calculated the COP at $\chi$-maximum figure of merit. The key of our proposal relies upon the state-like equation associated with the average $\langle x^2\rangle$, in a similar way as done in \cite{Contreras2023} for three stochastic heat engines. Our work suggest the construction of a Carnot-like refrigerator at microscopic level, in similar way as implemented by Blickle and Bechinger \cite{Blickle12} and Mart\'inez et al \cite{Martinez2015}, for stochastic heat engines.

\begin{acknowledgments}

OCV thanks the CONAHCyT-M\'exico for a scholarship. NSS thanks to SIP-IPN (M\'exico), as well as GVO acknowledges the support provided by CONAHCyT-M\'exico through the assignment postdoctoral fellowship: Estancias Posdoctorales por M\'exico 2022. 

\end{acknowledgments}

\bibliography{biblio}

\begin{thebibliography}{20}
\providecommand{\natexlab}[1]{#1}
\providecommand{\url}[1]{\texttt{#1}}
\expandafter\ifx\csname urlstyle\endcsname\relax
  \providecommand{\doi}[1]{doi: #1}\else
  \providecommand{\doi}{doi: \begingroup \urlstyle{rm}\Url}\fi

\bibitem[Contreras-Vergara et~al.(2023)Contreras-Vergara, S{\'a}nchez-Salas, Valencia-Ortega, and Jim{\'e}nez-Aquino]{Contreras2023}
O.~Contreras-Vergara, N.~S{\'a}nchez-Salas, G.~Valencia-Ortega, and J.~I. Jim{\'e}nez-Aquino.
\newblock Carnot, stirling, and ericsson stochastic heat engines: Efficiency at maximum power.
\newblock \emph{Physical Review E}, 108\penalty0 (1):\penalty0 014123, 2023.
\newblock \doi{doi.org/10.1103/PhysRevE.108.014123}.

\bibitem[Esposito et~al.(2010)Esposito, Kawai, Lindenberg, and Van~den Broeck]{Esposito10}
M.~Esposito, R.~Kawai, K.~Lindenberg, and C.~Van~den Broeck.
\newblock Efficiency at maximum power of low-dissipation carnot engines.
\newblock \emph{Phys. Rev. Lett.}, {\bf 105}\penalty0 (15):\penalty0 150603, 2010.
\newblock \doi{10.1103/PhysRevLett.105.150603}.

\bibitem[Schmiedl and Seifert(2007)]{Schmiedl2007}
T.~Schmiedl and U.~Seifert.
\newblock Efficiency at maximum power: An analytically solvable model for stochastic heat engines.
\newblock \emph{Europhysics Letters}, 81\penalty0 (2):\penalty0 20003, dec 2007.
\newblock \doi{10.1209/0295-5075/81/20003}.

\bibitem[De~Tom{\'a}s et~al.(2012)De~Tom{\'a}s, Hern{\'a}ndez, and Roco]{DeTomas12}
C.~De~Tom{\'a}s, A.~Calvo Hern{\'a}ndez, and J.~M.~M. Roco.
\newblock Optimal low symmetric dissipation carnot engines and refrigerators.
\newblock \emph{Phys. Rev. E}, 85\penalty0 (1):\penalty0 010104, 2012.
\newblock \doi{10.1103/PhysRevE.85.010104}.

\bibitem[Wang et~al.(2012)Wang, Li, Tu, Hern{\'a}ndez, and Roco]{Wang12}
Y.~Wang, M.~Li, Z.~C. Tu, A.~Calvo Hern{\'a}ndez, and J.~M.~M. Roco.
\newblock Coefficient of performance at maximum figure of merit and its bounds for low-dissipation carnot-like refrigerators.
\newblock \emph{Phys. Rev. E}, 86\penalty0 (1):\penalty0 011127, 2012.
\newblock \doi{10.1103/PhysRevE.86.011127}.

\bibitem[Apertet et~al.(2013)Apertet, Ouerdane, Goupil, and Lecoeur]{Apertet2013}
Y.~Apertet, H.~Ouerdane, C.~Goupil, and Ph. Lecoeur.
\newblock From local force-flux relationships to internal dissipations and their impact on heat engine performance: The illustrative case of a thermoelectric generator.
\newblock \emph{Phys. Rev. E}, 88:\penalty0 022137, Aug 2013.
\newblock \doi{10.1103/PhysRevE.88.022137}.

\bibitem[Holubec and Ryabov()]{Holubec2016}
V.~Holubec and A.~Ryabov.
\newblock Maximum efficiency of low-dissipation heat engines at arbitrary power.
\newblock \emph{J. Stat. Mech.: Theory Exp.}, 2016\penalty0 (7):\penalty0 073204.
\newblock \doi{10.1088/1742-5468/2016/07/073204}.

\bibitem[Gonzalez-Ayala et~al.(2016)Gonzalez-Ayala, Hern{\'a}ndez, and Roco]{Gonzalez2016}
J.~Gonzalez-Ayala, A.~Calvo Hern{\'a}ndez, and J.~M.~M. Roco.
\newblock Irreversible and endoreversible behaviors of the ld-model for heat devices: the role of the time constraints and symmetries on the performance at maximum $\chi$ figure of merit.
\newblock \emph{J. Stat. Mech.: Theory Exp.}, 2016\penalty0 (7):\penalty0 073202, jul 2016.
\newblock \doi{10.1088/1742-5468/2016/07/073202}.

\bibitem[Rana et~al.(2016)Rana, Pal, Saha, and Jayannavar]{Rana2016}
S.~Rana, P.S. Pal, A.~Saha, and A.M. Jayannavar.
\newblock Anomalous brownian refrigerator.
\newblock \emph{Phys. A: Stat. Mech. Appl.}, 444:\penalty0 783--798, 2016.
\newblock ISSN 0378-4371.
\newblock \doi{https://doi.org/10.1016/j.physa.2015.10.095}.

\bibitem[Ma et~al.(2018)Ma, Xu, Dong, and Sun]{Ma2018}
Y.~Ma, D.~Xu, H.~Dong, and C.~Sun.
\newblock Optimal operating protocol to achieve efficiency at maximum power of heat engines.
\newblock \emph{Phys. Rev. E}, 98:\penalty0 022133, Aug 2018.
\newblock \doi{10.1103/PhysRevE.98.022133}.

\bibitem[Johal(2019)]{Johal2019}
R.~S. Johal.
\newblock Performance optimization of low-dissipation thermal machines revisited.
\newblock \emph{Phys. Rev. E}, 100:\penalty0 052101, Nov 2019.
\newblock \doi{10.1103/PhysRevE.100.052101}.

\bibitem[Ma et~al.(2021)Ma, Sun, and D.]{Ma2021}
Y.~Ma, C.~P. Sun, and Hui D.
\newblock Consistency of optimizing finite-time carnot engines with the low-dissipation model in the two-level atomic heat engine.
\newblock \emph{Commun. Theor. Phys.}, 73\penalty0 (12):\penalty0 125101, nov 2021.
\newblock \doi{10.1088/1572-9494/ac2cb8}.

\bibitem[Chen et~al.(2022)Chen, Chen, Fei, and Quan]{Chen2022}
Y.~H. Chen, Jin-Fu Chen, Zhaoyu Fei, and H.~T. Quan.
\newblock Microscopic theory of the curzon-ahlborn heat engine based on a brownian particle.
\newblock \emph{Phys. Rev. E}, 106:\penalty0 024105, Aug 2022.
\newblock \doi{10.1103/PhysRevE.106.024105}.

\bibitem[Zhao et~al.(2022)Zhao, Gong, and Tu]{Zhao2022}
Xiu-Hua Zhao, Zheng-Nan Gong, and Z.~C. Tu.
\newblock Low-dissipation engines: Microscopic construction via shortcuts to adiabaticity and isothermality, the optimal relation between power and efficiency.
\newblock \emph{Phys. Rev. E}, 106:\penalty0 064117, Dec 2022.
\newblock \doi{10.1103/PhysRevE.106.064117}.

\bibitem[Curzon and Ahlborn(1975)]{Curzon1975}
F.~L. Curzon and B.~Ahlborn.
\newblock Efficiency of a carnot engine at maximum power output.
\newblock \emph{Am. J. Phys.}, {\bf 43}\penalty0 (1):\penalty0 22--24, 1975.
\newblock \doi{10.1119/1.10023}.

\bibitem[Yan and Chen(1990)]{Yan90}
Z.~Yan and J.~Chen.
\newblock A class of irreversible carnot refrigeration cycles with a general heat transfer law.
\newblock \emph{Journal of Physics D: Applied Physics}, 23\penalty0 (2):\penalty0 136, 1990.
\newblock \doi{10.1088/0022-3727/23/2/002}.

\bibitem[Blickle and Bechinger(2012)]{Blickle12}
V.~Blickle and C.~Bechinger.
\newblock Realization of a micrometre-sized stochastic heat engine.
\newblock \emph{Nat. Phys.}, {\bf 8}\penalty0 (2):\penalty0 143--146, 2012.
\newblock \doi{10.1038/nphys2163}.

\bibitem[Mart{\'\i}nez et~al.(2016)Mart{\'\i}nez, Rold{\'a}n, Dinis, Petrov, Parrondo, and Rica]{Martinez16}
I.~A. Mart{\'\i}nez, {\'E}.~Rold{\'a}n, L.~Dinis, D.~Petrov, J.~M. Parrondo, and R.~A. Rica.
\newblock Brownian carnot engine.
\newblock \emph{Nat. Phys.}, {\bf12}:\penalty0 67--70, 2016.
\newblock \doi{10.1038/nphys3518}.

\bibitem[Mart{\'\i}nez et~al.(2015)Mart{\'\i}nez, Rold{\'a}n, Dinis, Petrov, and Rica]{Martinez2015}
I.~A. Mart{\'\i}nez, {\'E}.~Rold{\'a}n, L.~Dinis, D.~Petrov, and R.~A. Rica.
\newblock Adiabatic processes realized with a trapped brownian particle.
\newblock \emph{Phys. Rev. Lett.}, {\bf 114}\penalty0 (12):\penalty0 120601, 2015.
\newblock \doi{10.1103/PhysRevLett.114.120601}.

\bibitem[Qi et~al.(2022)Qi, Chen, Ge, Huang, and Feng]{qi2022}
C.~Qi, L.~Chen, Y.~Ge, L.~Huang, and H.~Feng.
\newblock Thermal brownian refrigerator with external and internal irreversibilities.
\newblock \emph{Case Stud. Therm. Eng.}, 36:\penalty0 102185, 2022.
\newblock \doi{10.1016/j.csite.2022.102185}.

\end{thebibliography}
\vskip3.0cm

\end{document}